\documentclass[reprint,twocolumn,showpacs,preprintnumbers,superscriptaddress,amsmath,amssymb]{revtex4-2}
\usepackage{graphicx}
\usepackage{siunitx}
\usepackage{xcolor}
\usepackage{amsmath,amssymb}
\usepackage[T1]{fontenc}
\usepackage{lmodern}
\usepackage[utf8]{inputenc}
\usepackage[english]{babel}
\usepackage{natbib}
\usepackage{hyperref}
\usepackage{mathtools}
\hypersetup{colorlinks=true,citecolor={blue},linkcolor={blue},urlcolor={blue}}
\usepackage{ulem}

\begin{document}

\title{
Shaping Bulk Fermi Arcs in the Momentum Space of Photonic Crystal Slabs}
\author{Luigi Frau}
\email{l.frau02@gmail.com}
\affiliation{
Dipartimento di Fisica ``A. Volta,'' Universit\`a di Pavia, via Bassi 6, 27100 Pavia (Italy)
}
\affiliation{CNR Nanotec, Institute of Nanotechnology, via Monteroni, 73100, Lecce (Italy)}
\author{Simone Zanotti}
\affiliation{
Dipartimento di Fisica ``A. Volta,'' Universit\`a di Pavia, via Bassi 6, 27100 Pavia (Italy)
}
\author{Lydie Ferrier}%
\affiliation{Ecole Centrale de Lyon, CNRS, INSA Lyon, \\ Universit\'e  Claude Bernard Lyon 1, CPE Lyon, CNRS, INL, UMR5270, Ecully 69130 (France)}
\author{Dario Gerace}%
\affiliation{
Dipartimento di Fisica ``A. Volta,'' Universit\`a di Pavia, via Bassi 6, 27100 Pavia (Italy)
}
\affiliation{CNR Nanotec, Institute of Nanotechnology, via Monteroni, 73100, Lecce (Italy)}
\author{Hai Son Nguyen}
\email{hai-son.nguyen@ec-lyon.fr}
\affiliation{Ecole Centrale de Lyon, CNRS, INSA Lyon, \\ Universit\'e  Claude Bernard Lyon 1, CPE Lyon, CNRS, INL, UMR5270, Ecully 69130 (France)}
\affiliation{Institut Universitaire de France (IUF), 75231 Paris (France)}

\date{\today}

\begin{abstract}
Exceptional points (EPs) are special spectral degeneracies of non-Hermitian operators: at the EP, the complex eigenvalues coalesce, i.e., they become degenerate in both their real and imaginary parts. In two-dimensional (2D) photonic crystal lattices, these elements can be tailored through structural engineering. In particular, it is known that a quadratic degeneracy in the photonic band structure can be split into a pair of Dirac points (DPs) by breaking one of the unit cell symmetries, and each DP can be further split into a pair of EPs by introducing losses. Each EP of the pair is then connected by an open isofrequency curve, called the bulk Fermi arc (BFA). In this work, we introduce a simplified effective Hamiltonian model accounting for the main physical properties of these EPs and BFAs. Then, we systematically investigate, through numerical simulations, how EPs as well as the related BFA depend on the type and amount of broken symmetries in the given 2D unit cell of a realistic photonic crystal slab implementation. Our results show that it is possible to tailor the position and distance of the EP pair in reciprocal space, as well as the curvature and orientation of the associated BFA, by deterministically tuning the unit cell structure. Importantly, the symmetry-breaking strategy we propose is general and can be applied to a broad range of photonic crystal designs beyond the specific example studied here. This approach opens new possibilities for exploiting EPs in applications involving photonic crystal lattices in, e.g., light-emitting devices or fundamental physics studies.
\end{abstract}

\maketitle


\section{\label{intro} Introduction}
Over the past decade, non-Hermitian physics has witnessed remarkable developments within photonic platforms, encompassing both purely lossy systems and parity-time (PT) symmetric systems that exhibit a balance between optical gain and losses \cite{El-Ganainy2019,Yan2023,Nasari2023,Li2023}. These advances have been made possible by the ability to engineer synthetic optical materials at subwavelength scales, thus enabling precise control over their gain and loss properties. The physics of such open systems can be described by using effective non-Hermitian Hamiltonian models \cite{rotter2009}, characterized by complex eigenvalues and non-orthogonal eigenvectors. Within this context, of particular interest are the degeneracies of non-Hermitian Hamiltonians, known as Exceptional Points (EPs), characterized by the coalescence of both eigenvalues and eigenvectors. EPs represent topological singularities in the non-Hermitian parameter space \cite{Kawabata2019,Bergholtz2021,okuma2023,ding2022,Nasari2022,Li2023}, and their exotic properties have enabled the demonstration of a variety of non-trivial physical phenomena, including mode coalescence in coupled waveguides \cite{benisty2015,Goldzak2018,Khurgin2021,Schumer2022}, unidirectional and reflectionless light propagation \cite{Lin2011,Regensburger2012}, loss-gain-enabled microresonator lasing \cite{Peng2014,peng2014b,brandstetter2014,feng2014,hodaei2014,Shahmohammadi2022}, enhanced sensing sensitivity \cite{chen2017,hodaei2017,park2020}, increased local density of optical states and spontaneous emission \cite{Pick2017,Takata2021,ferrier2022}, anomalous decay dynamics \cite{Wu2020}, exceptional bound states in the continuum (BICs) \cite{Minkov2018,Valero_ExceptionalBICs2025}, and geometry-dependent skin effects\cite{Fang2022}.\\

In general, a pair of EPs can emerge from an isolated Dirac point (DP) upon the introduction of non-Hermitian perturbations. These EPs carry opposite half-quantized topological charges, which originate from the eigenvalue braiding around each EP. As a consequence, the EP pair is connected by a bulk Fermi arc (BFA), i.e., an open isofrequency curve of purely real degeneracy~\cite{Zhou2018,Kawabata2019,Kozii2024}. It is important to distinguish this BFA — a non-Hermitian feature of the bulk dispersion in two-dimensional (2D) systems — from the Fermi arc surface states observed in 3D Weyl semimetals \cite{Wan2011,Noh2017,Nguyen2023}. Notably, the BFA is topologically robust due to the definition of charges of the EPs as topological invariants, and it can only vanish when the EPs merge and annihilate \cite{Krl2022}. To date, BFAs have been experimentally observed in photonic systems in which pairs of EPs arise from symmetry-protected Dirac points \cite{Zhou2018,Su2021,Krl2022}. However, most of the existing studies focus solely on locating EPs, and there has been little exploration on how to engineer the dispersion characteristics of the BFA itself — such as its shape, curvature, and orientation in momentum space — as a function of the photonic lattice  parameters.

In this work, we discuss a realistic photonic crystal implementation allowing to deterministically tune the symmetry properties with the aim of engineering the BFA in reciprocal space. We introduce a simplified two-band model that captures all the essential features of this phenomenon, based on a few adjustable parameters. Then we systematically investigate, through extensive numerical simulations, how the positions of EPs and the properties of the associated BFAs depend on the type and degree of symmetry breaking in the 2D unit cell of a realistic photonic crystal implementation. Our numerical study focuses on 2D lattices implemented in thin dielectric films (2D photonic crystal slabs), and reveals that it is possible to precisely control both the location and separation of EP pairs in reciprocal space, as well as the curvature and orientation of the connecting BFA, by only controlling two structural parameters in the lattice design. The proposed symmetry-based strategy for tailoring EPs and BFAs is general and can be extended to a broad class of photonic crystal architectures beyond the specific examples presented here. These findings open up new avenues for the application of EPs in photonic crystal-based devices, including light-emitting structures and platforms for exploring fundamental aspects of non-Hermitian physics.

\section{Theory and methods}

\subsection{\label{model} Photonic crystal slab design}
The photonic platform investigated in this work is a partially etched photonic crystal (PhC) slab, composed of a thin silicon nitride (SiN) film—serving as the core layer of a planar waveguide patterned atop a uniform silica (SiO$_2$) substrate, which acts as the lower cladding. This material configuration is widely used in integrated Photonics, and it has been adopted in numerous experimental studies of guided resonances and non-Hermitian photonic phenomena, owing to its low optical loss, high index contrast, and compatibility with standard nanofabrication techniques\,\cite{Zhou2018,Nguyen2023EP,Yin2025}.

It is well established that the mode spectrum of PhC slabs includes: fully guided modes, which lie outside the light cones of all cladding layers and are ideally lossless; quasi-guided modes, or guided resonances, which are largely confined within the waveguide core but experience radiative leakage due to coupling with continuum modes inside the light cone; and fully radiative modes, which are completely delocalized and radiate electromagnetic energy outside the waveguide plane\,~\cite{Sakoda2001,Fan2002,Tikhodeev2002,Gerace2004,Andreani2006}. Among these, guided resonances are of particular interest here, since they can be engineered to exhibit complex eigenfrequencies with both real and imaginary components, i.e., key features in view of controlling the dispersion of EPs and the associated BFAs.

\begin{figure}
    \centering
    \includegraphics[width=0.8
    \linewidth]{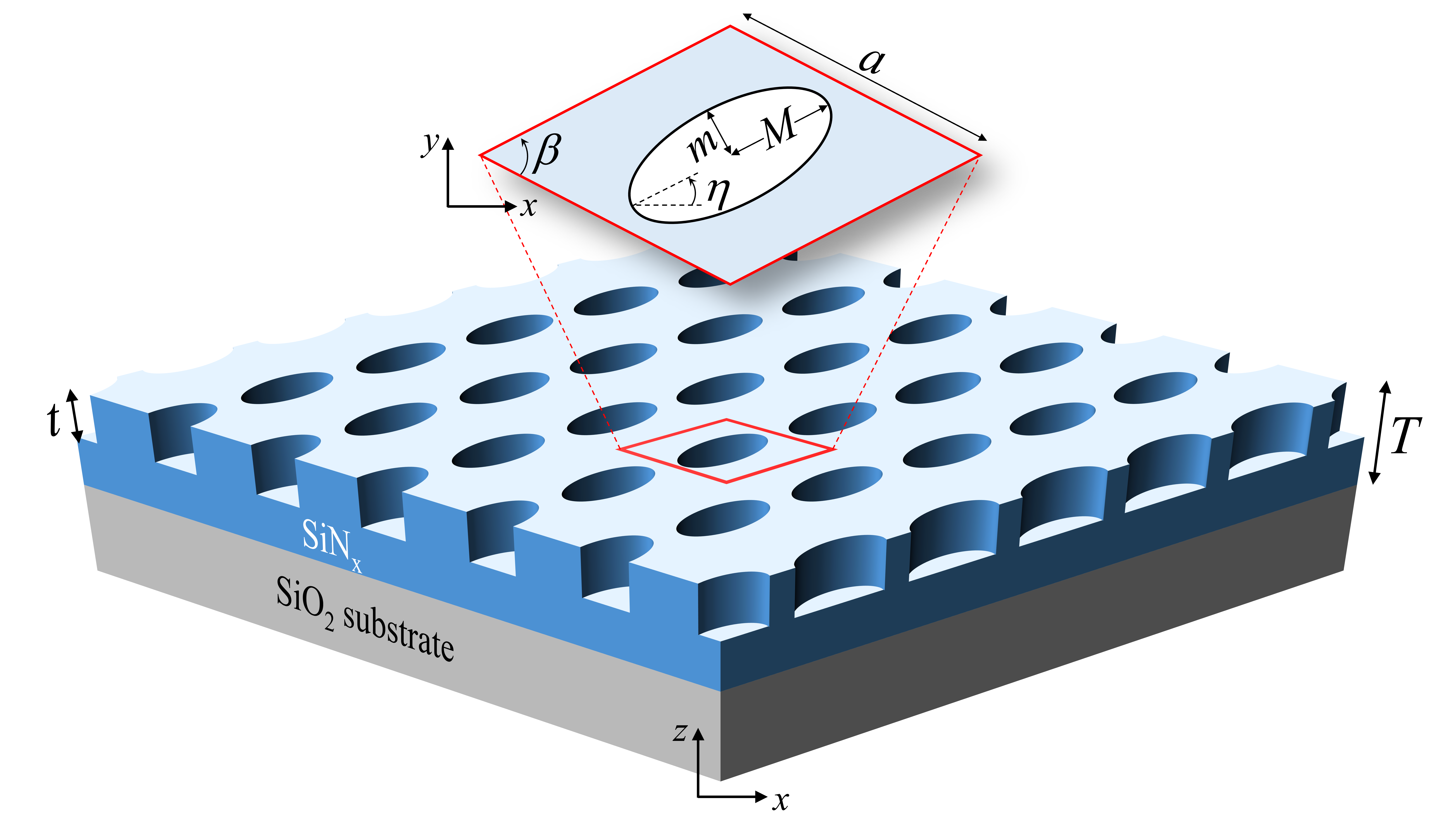}
    \caption{Schematic illustration of the basic building blocks defining the photonic crystal lattice parameters studied in this work: the rhombic unit cell, defined by a deformation angle $\beta$, and the cross section of the planar dielectric waveguide patterned with a 2D photonic lattice partially etched with depth $t$ into the waveguide core with total thickness $T$. The zoom shows the relevant hole parameters considered in this work, namely an elliptical shape defined by its  short ($m$) and long ($M$) semi-axes, and by the rotation angle with respect to the horizontal axis, $\eta$.
    }
    \label{fig:system}
\end{figure}

With the aim of studying the formation and evolution of EPs in such a system, we remind that the process begins with the emergence of crossing points between photonic bands of opposite in-plane symmetry along some specific symmetry direction in reciprocal space. These degeneracies, analogous to DPs in lossless systems, arise from intentional distortions of the lattice unit cell, as it has been already discussed in the literature\,\cite{Chong2008}. When in-plane symmetry is broken, the two photonic bands can couple, and their mutual interaction — combined with differing radiative loss rates — leads to the formation of EP pairs emanating from each crossing\,\cite{Zhou2018,Fang2022}. The required symmetry breaking can be introduced either by operating at non-zero in-plane wavevectors within a mirror plane, or by geometrically modifying the unit cell motif.
Here we explore a PhC design in which the unit cell of the 2D photonic lattice is a rhombus containing an elliptical air hole, as schematically shown in Fig.~\ref{fig:system}. The lattice periodicity, $a$, is defined as the side length of the rhombus, with primitive vectors given as
\begin{align}
&\vec{a}_1 = a\cos\left(\frac{\beta}{2}\right){\hat{\mathbf{x}}} + a\sin\left(\frac{\beta}{2}\right){\hat{\mathbf{y}}} \nonumber \\
&\vec{a}_2 = -a\cos\left(\frac{\beta}{2}\right){\hat{\mathbf{x}}} + a\sin\left(\frac{\beta}{2}\right){\hat{\mathbf{y}}}    \, ,
\end{align}
in which $\hat{\mathbf{x}}$ and $\hat{\mathbf{y}}$ define the Cartesian unit vectors, and $\beta \in (0,90^\circ]$ is the internal angle. \\
When $\beta = 90^\circ$ and the hole is circular, the structure possesses full $C_{4v}$ symmetry, i.e., it coincides with a square lattice of circular holes.
By reducing $\beta$ below $90^\circ$, the lattice symmetry is lowered to $C_{2v}$, preserving only two mirror operations: $\sigma_x$ (i.e., mirror reflection across the $yz$ plane, or transformations such that $x \rightarrow -x$), and $\sigma_y$ (mirror reflection across the $xz$ plane, i.e., $y \rightarrow -y$). In the following, these symmetries will be selectively broken: $\sigma_x$ is broken, e.g., by working at $k_x \neq 0$, while $\sigma_y$ can be broken by working at $k_y \neq 0$. Additionally, if $\eta$ denotes the angle between the major axis of the ellipse and the horizontal direction (coinciding with the major rhombus axis), as represented in Fig.~\ref{fig:system}, $\sigma_y$ is also broken at $k_y = 0$ by setting $\eta \neq 0^\circ$.

\subsection{\label{sec:concept}
Analytic model: emergence of EPs by mode coupling via symmetry breaking}

\begin{figure}[t!]
    \centering
    \includegraphics[width=0.6\linewidth]{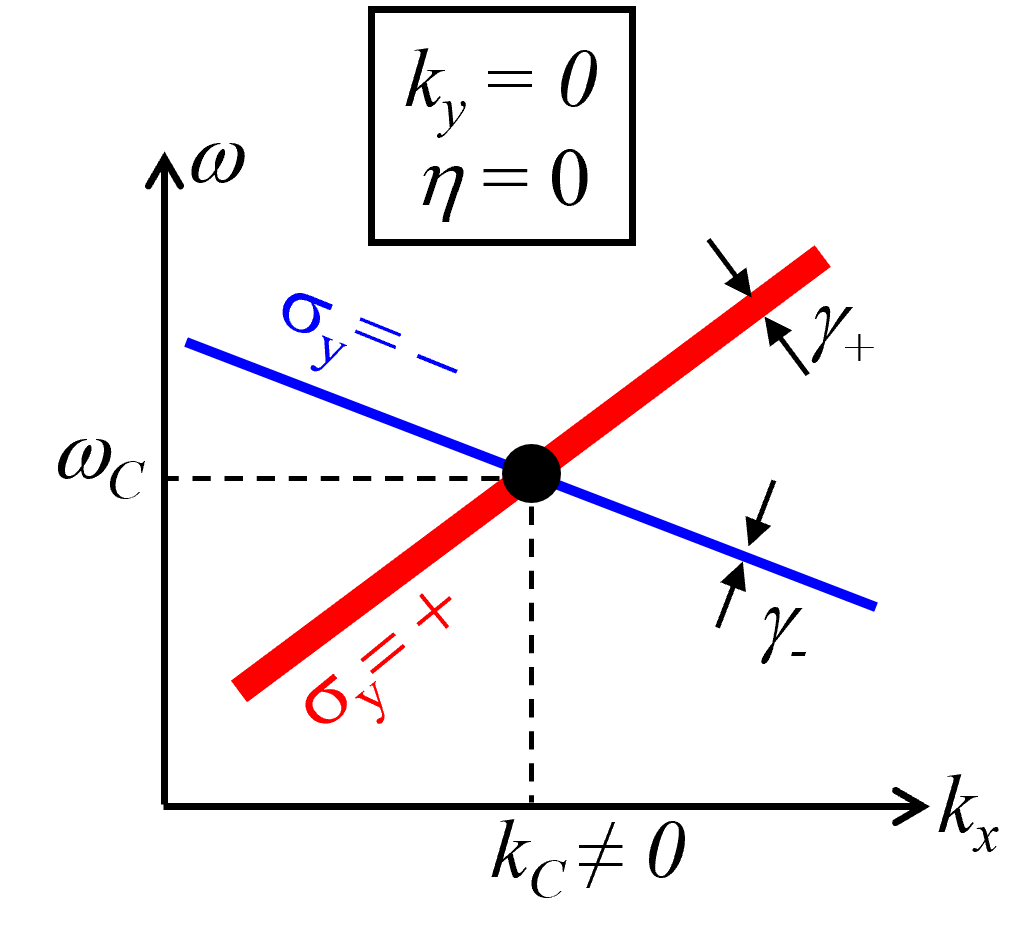}
    \caption{Schematic illustration of the crossing between two photonic bands with linear $k_x$ dispersion possessing different intrinsic losses (represented by linewidths $\gamma_{\pm}$), and exhibiting opposite $\sigma_y$ symmetry, crossing at $k_x=k_C$ and $k_y=0$ in reciprocal space since no $\sigma_y$ symmetry is broken. }
    \label{fig:crossing_sketch}
\end{figure}

The general mechanism for the emergence of EPs in our system is the hybridization of two modes with opposite $\sigma_y$ symmetry. As already mentioned at the end of the last Section, this hybridization occurs when $\sigma_y$ symmetry is broken either by working at $k_y \neq 0$ or by a geometric tilting angle of the elliptical hole, $\eta \in (0^\circ, 90^\circ)$. These EPs appear near the crossing points of two photonic bands that differ both in radiative losses and in their $\sigma_y$ parity, i.e., one being even (eigenvalue $\sigma_y = +1$) and the other odd ($\sigma_y = -1$) against mirror reflection. As sketched in Fig.~\ref{fig:crossing_sketch}, such a mode crossing generally occurs at $k_x\neq 0$, where the $\sigma_x$ symmetry is already broken, while requiring $k_y = 0$ and $\eta = 0$ to maintain well-defined $\sigma_y$ symmetry. In the following, we detail how EPs arise when this symmetry is broken by $k_y \neq 0$ and/or $\eta\neq 0$, respectively.

\subsubsection{The $\eta = 0^{\circ}$ case}
We first consider the case in which the unit cell still preserves $C_{2v}$ symmetry (i.e., null $\eta$). Assuming the band crossing occurs at $(k_x, k_y) = (k_C, 0)$ with energy $\omega = \omega_C$, the complex mode detuning between the two uncoupled modes can be expressed in a simple analytic form as:
\begin{equation}
\omega_+(k_x,k_y) - \omega_-(k_x,k_y) = v(k_x - k_C) + \zeta k_y^2 + i\gamma \, .
\end{equation}
Here, the term $v(k_x - k_C)$ arises from the linear dispersion of the two bands around the crossing point along $k_x$, as schematically depicted in Fig.~\ref{fig:crossing_sketch}, while the quadratic term, $\zeta k_y^2$, reflects the even parity of the modes with respect to $k_y$ as imposed by the preserved $\sigma_y$ symmetry of the unit-cell design. For simplicity, the imaginary part, $\gamma = \gamma_+ - \gamma_-$,  accounting for the difference in radiative losses between the two modes, is assumed to be constant in the vicinity of the crossing point.

When $k_y \neq 0$, the $\sigma_y$ symmetry is broken, thus allowing the two modes to be coupled with a strength proportional to $k_y$. The simplest model capturing the effects of this coupling can be represented as an effective non-Hermitian Hamiltonian (after removing the global energy and loss offset):
\begin{equation}
H = \begin{pmatrix}
\frac{v(k_x - k_C) + \zeta k_y^2 + i\gamma}{2} && \alpha k_y \\\
\alpha^* k_y && -\frac{v(k_x - k_C) + \zeta k_y^2 + i\gamma}{2} 
\end{pmatrix} \, ,
\end{equation}
in which the off-diagonal term, $\alpha k_y$, 
mimics an effective spin-orbit-like coupling between the two modes, thus linking their symmetry (pseudospin) to momentum. While this Hamiltonian term should not be considered, strictly speaking, as a true spin-orbit interaction, it plays an analogous role in shaping the dispersion and enabling non-Hermitian degeneracies. Here, $\alpha$ can be taken to be real without loss of generality. After diagonalization, we get the two complex eigenenergies.
Hence, by evaluating the complex energy gap, $\Delta\omega=\{\left[v(k_x - k_C) + \zeta k_y^2 + i\gamma\right]^2 + 4(\alpha k_y)^2\}^{1/2}$, two EPs are obtained, corresponding to $\Delta\omega=0$. After the analytic solution, we find that these two points are localized in momentum space at coordinates:
\begin{equation}
\begin{split}
k_x^{\text{EP}_{1,2}} &= k_C - \frac{\zeta \gamma^2}{4v \alpha^2} \, , \\
k_y^{\text{EP}_{1,2}} &= \pm \frac{\gamma}{2\alpha} \, .
\end{split}
\end{equation}
These two EPs are symmetrically positioned on opposite sides of the $k_y = 0$ axis. Their mutual distance along the $k_y$ direction can be calculated as
\begin{equation}
\Delta k_y = |k_y^{\text{EP}_1} - k_y^{\text{EP}_2}| = {|\gamma|}/{|\alpha|}. \label{eq:eta0_Deltaky}
\end{equation}
Thus, $\Delta k_y $ can be tuned by adjusting the ratio ${|\gamma|}/{|\alpha|}$. In our photonic crystal design, this is practically achieved by varying the internal angle $\beta$ of the rhombic unit cell, as it will be shown in the following.

Moreover, this simplified analytic model allows to straightforwardly define the BFA connecting the two EPs, which satisfies the parabolic dispersion in reciprocal space:
\begin{equation}
k_x = k_C + \mu k_y^2 \, , \label{eq:eta0_BFA}
\end{equation}
in which  $\mu=-{\zeta}/{v}$ defines the Fermi arc curvature, clearly governed by the ratio ${\zeta}/{v}$. As it will be shown in the following, this curvature can be continuously tuned by changing $\beta$ in our photonic lattice designs, thus allowing for positive, zero, or negative curvature (see numerical results in Sec.~\ref{sec:results}).

\subsubsection{The $\eta \neq 0^{\circ}$ case}
Introducing a non-zero $\eta$ breaks the $C_2v$ symmetry of the unit cell, and it has two distinct effects. First, the even parity of the band dispersion with respect to $k_y$ — which still holds at $k_x\neq 0$ when $\sigma_y$ is preserved — is no longer satisfied. This asymmetry introduces a linear term into the detuning between the uncoupled modes, which we can assume to be of the form $\kappa \eta k_y$, and it also modifies the coefficient of the quadratic term as $\zeta(\eta) k_y^2$, i.e., with $\zeta$ that is now explicitly dependent on $\eta$. Second, the symmetry breaking gives rise to an additional coupling term proportional to $\eta$. As a result, the effective non-Hermitian Hamiltonian becomes:
\begin{equation}\label{eq:analytic_eta_nonzeero}
H' = \begin{pmatrix}
\frac{v(k_x - k_C) +\kappa\eta k_y+\zeta k_y^2 + i\gamma}{2} && \alpha k_y + \delta\eta\\\
\alpha^* k_y + \delta^*\eta&& -\frac{v(k_x - k_C) + \kappa\eta k_y + \zeta k_y^2 + i\gamma}{2}
\end{pmatrix} \, .
\end{equation}
While a global phase rotation of the basis can always bring one of the coupling terms to be real, it is not possible, in general, to make both $\alpha$ and $\delta$ simultaneously real unless they share the same complex phase. However, for simplicity of the model — which still captures the essential physics of the systems under investigation — we assume both $\alpha$ and $\delta$ to be real, in the following.  By evaluating now the complex gap from the two eigenvalues of Eq.~\eqref{eq:analytic_eta_nonzeero}, we obtain $\Delta\omega=\{\left[v(k_x - k_C) + \kappa\eta k_y + \zeta k_y^2 + i\gamma\right]^2 + 4(\alpha k_y+\delta\eta)^2\}^{1/2}$, from which the two EPs fulfilling the condition $\Delta\omega=0$ are now located at:
\begin{equation}
\begin{split}
k_x^{\text{EP}_{1,2}} &= k_C - \frac{\kappa\eta}{v}k_y^{\text{EP}_{1,2}} -\frac{\zeta}{v} \left(k_y^{\text{EP}_{1,2}}\right)^2 \, , \\
k_y^{\text{EP}_{1,2}} &= \frac{\pm\gamma-\delta\eta}{2\alpha}\, .
\end{split}
\end{equation}
When $\eta \neq 0$, the two EPs are no longer symmetrically located about the $k_y = 0$ axis. In particular, if $|\eta| > {|\gamma|}/{|\delta|}$, both EPs reside in the same half of the momentum space. To quantify the effect of $\eta$ on the EP configuration, we define the vertical displacement of the EP pair from the $k_y = 0$ line by the midpoint:
\begin{equation}
k_{y,0} = \frac{1}{2} \left(k_y^{\text{EP}_1} + k_y^{\text{EP}_2}\right) = -\frac{\delta}{\alpha}\eta \, ,
\end{equation}
and the angle $\theta$ between the line connecting the two EPs and the $k_x$-axis by:
\begin{equation}
\tan\theta = \frac{k_y^{\text{EP}_1} - k_y^{\text{EP}_2}}{k_x^{\text{EP}_1} - k_x^{\text{EP}_2} }=\frac{v}{\frac{\zeta\delta}{\alpha} - \kappa} \cdot \frac{1}{\eta} \, .
\end{equation}
These expressions show that the location of the EP pair and their relative orientation in momentum space can be controlled by simply tuning $\eta$.

Finally, the BFA connecting the two EPs is described by:
\begin{equation}
k_x = k_C + \rho k_y + \mu k_y^2\, ,
\end{equation}
with $\rho=-{\kappa\eta}/{v}$ and $\mu=-{\zeta}/{v}$.
Here, we note that $\zeta$ may itself depend on $\eta$. As such, varying $\eta$ not only shifts and tilts the BFA (through the linear term of slope, $\rho$) but may also invert its curvature $\mu$, thus enabling a transition from a concave to a convex profile — or even a flat one. This tunability provides a powerful degree of control over non-Hermitian band topology via geometric design.

In summary, despite its simplicity our effective two-mode model captures the essential physics underlying the tailoring of EPs and BFAs in momentum space. In the following Section, we are going to quantitatively validate the model basic predictions through rigorous numerical simulations of photonic modes in our proposed structured design, demonstrating the controllable emergence of EPs and BFAs in both $C_{2v}$ symmetric (i.e., for $\eta = 0^{\circ}$) and symmetry-broken ($\eta \neq 0^{\circ}$) configurations.

\subsection{Numerical Methods}
To investigate the photonic properties, including EPs and the associated BFAs, resulting from the variations of structural parameters in the proposed PhC slab structure, Maxwell equations are solved numerically for each realization of the system under investigation. In particular, two complementary approaches are employed: rigorous coupled-wave analysis (RCWA)\,\cite{Whittaker1999,Liscidini2008}, and the finite element method (FEM)\,\cite{Jin_FEM_book}. For the numerical implementation of RCWA, we use the open-source package $S^4$,\cite{s4}, which computes scattering matrices under plane-wave excitation, thus allowing to calculate spectral quantities such as reflection, transmission, and absorption for both transverse electric (TE) and transverse magnetic (TM) polarizations, thus providing insight into the resonance behavior of the structure at varying angles. In the $S^4$ simulations, the dielectric materials under investigation have been assumed with refractive indices 
$n_{\mathrm{SiO}_2}=1.46$ and $n_{\mathrm{SiN}}=1.954$ for the lower cladding and core layers, respectively. In addition, a small imaginary part of the SiN refractive index (i.e., \textbf{$10^{-8}$}) is assumed in order to better visualize absorption peaks. In all the photonic designs provided in the following, the total thickness of the PhC slab (i.e., the core layer thickness) is fixed as $T=0.5a$, while the etching depth is given as $t=0.25a$ (see schematic defintion of these parameters in Fig.~\ref{fig:system}).

\begin{figure*}[t]
    \centering
    \includegraphics[width=0.9\textwidth]{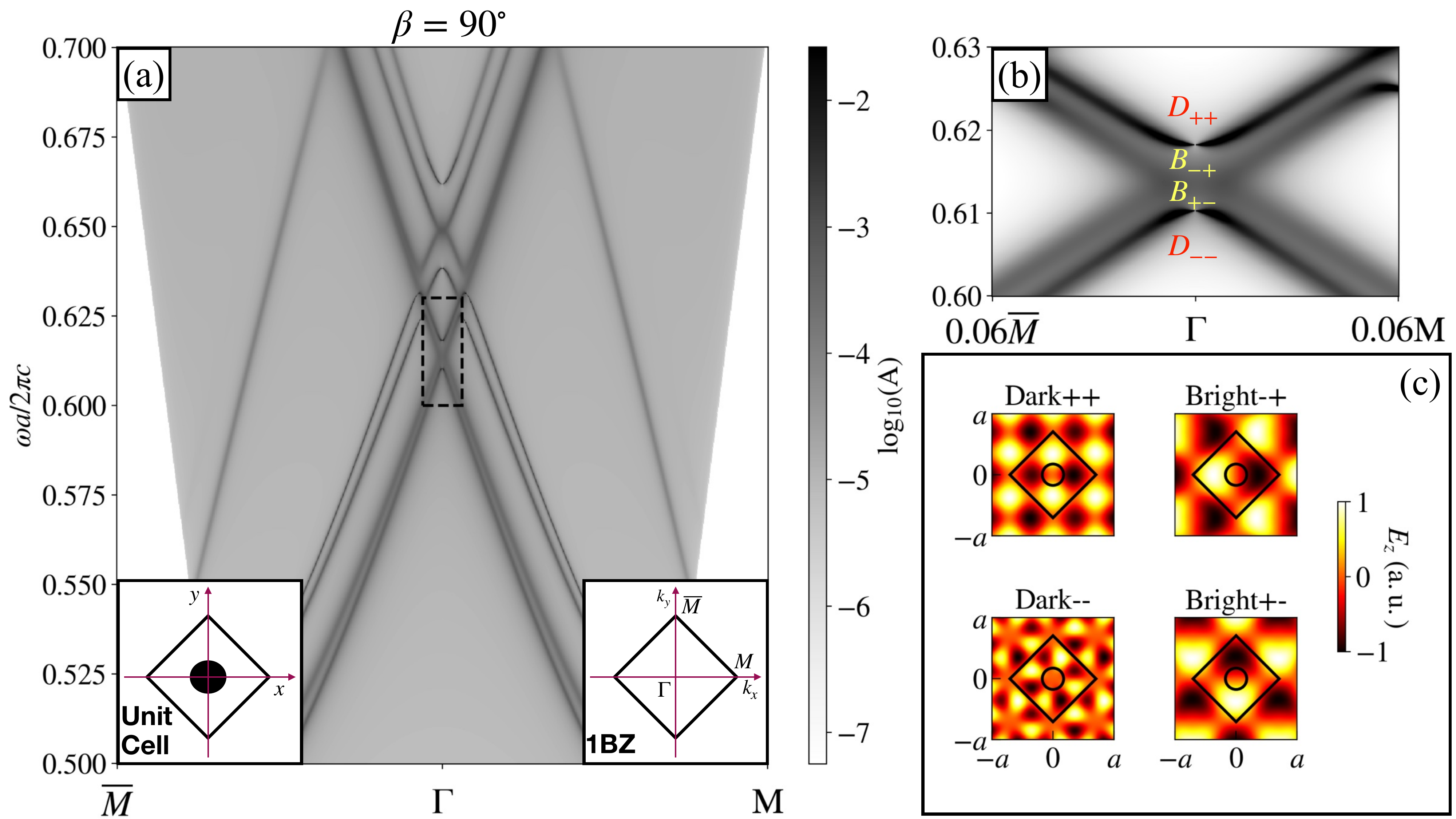}
    \caption{Dispersion within the light cone of the photonic resonances along the $\Gamma$M direction within the first Brillouin zone of the square lattice (see insets), as obtained from the numerically simulated optical response (by RCWA): quasi-guided modes are clearly visible as peaks in the variable angle absorption spectra (plotted in log scale, log$_{10}(A)$, dark areas being associated to larger absorption) calculated for the partially etched SiN PhC slab with (a) the undistorted lattice unit cell (corresponding to $\beta=90^{\circ}$) and a circular hole at its center (see inset). A close up of the dispersion around $\Gamma$ is reported in (b), corresponding to the dashed rectangular area in (a), showing the presence of two dark modes (upper and lower) and two degenerate bright ones at $\Gamma$. The corresponding field profiles ($z$-component of the electric field) are plotted in (c) for the four relevant modes of panel (b), see labels indicating the respective $\sigma_x$ and $\sigma_y$ symmetries at $\Gamma$. In the simulations, the circular hole radius is set to $m/a=0.34$ (with $m=M$ in this case),
    and all the energies are then given in dimensionless units, $a/\lambda=\omega a/2\pi c$. }
    \label{fig:dispersions_S4}
\end{figure*}

As an alternative approach to the calculation of the optical response by $S^4$, we also employ the the commercial software COMSOL Multiphysics (hereafter referred to as COMSOL)\,\cite{comsol}, which allows for a direct computation of the complex eigenfrequencies through the numerical implementation of FEM. This method discretizes the structure in real space with a triangular mesh, and then solves the eigenvalue problem for the stationary (i.e., time independent) Maxwell operator, thus enabling the extraction of both real and imaginary parts of the resonant modes. In the COMSOL implementation, a single unit cell is set up in the simulation layout, and Floquet boundary conditions are imposed along the plane boundaries to simulate an infinitely and periodically extended 2D PhC lattice. Absorbing boundary conditions are set up at the top and bottom of the structure, in order to simulate semi-infinitely extended upper and lower claddings. This implementation is obtained with perfectly matched layers (PML) boundary conditions, which efficiently extinguish vertically outgoing waves. Convergence of the results as a function of the triangular mesh has been checked, and a maximum mesh size of 20\% of the vacuum wavelength has been used in all the simulations.


\begin{figure*}[t]
    \centering
    \includegraphics[width=0.9\textwidth]{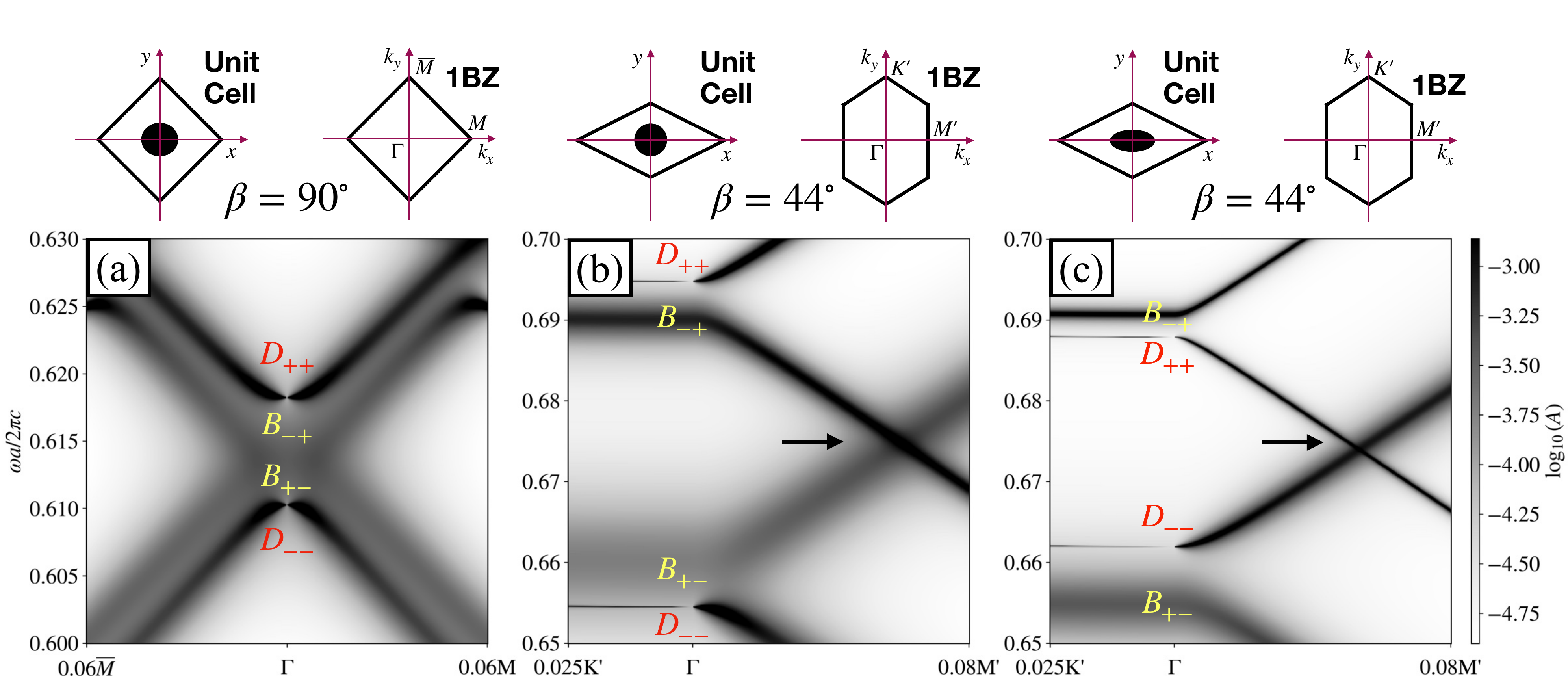}
    \caption{Close up of the photonic mode dispersion within the light cone in the vicinity of the $\Gamma$ point, and along two orthogonal directions ($x$ and $y$), for the lattice structure reported in the figures. Numerical simulations of absorption spectra have been obtained by RCWA: quasi-guided modes are clearly visible as peaks in the variable angle absorption spectra (log scale, log$_{10}(A)$, dark areas associated to larger absorption), calculated for the same partially etched SiN slab as in the previous Figure. In panel (a) we report the same results as in Fig.~\ref{fig:dispersions_S4}(b), for direct comparison, which corresponds to the undistorted lattice unit cell ($\beta=90^{\circ}$); in panel (b) we report results for the distorted unit cell (i.e., $\beta=44^{\circ}$ here) with a circular hole at its center, while in (c) we report the results for the same distorted cell but including an elliptical hole at its center (see top schemes), from which the evolution from the quadratic degeneracy point at $\Gamma$ (see panel a) to the Dirac crossing at $k_x\neq 0$ (panels b and c) can be appreciated. Mode symmetries at $\Gamma$ are explicitly indicated. In panel (c), the Dirac point corresponds to the crossing of D$_{++}$ and D$_{--}$ modes, respectively. Photonic crystal parameters are as in Fig.~\ref{fig:dispersions_S4}, and all the energies are then given in dimensionless units, $a/\lambda=\omega a/2\pi c$. In (b) the hole radius is set to $m/a=0.28$ (with $m=M$), while in (c) we set $m/a=0.2$ and $M/a=0.4$. With this choice, the air fraction is kept constant from (a) to (c).}
    \label{fig:dispersions_S4_zoom}
\end{figure*}

\section{\label{results} Numerical Results}

We hereby report the results of our numerical simulations on the model system represented in  Fig.~\ref{fig:system}, which has been first analysed by RCWA to investigate the broadband photonic mode dispersion. Then, we will report FEM results to specifically focus on the narrow band properties around the DPs and EPs, which would be difficult to capture by RCWA only. The numerical study has focused on the dependence of the complex energy dispersion of guided resonances supported by the PhC slab from structural parameters of the unit cell. In particular, scanning the two angles, $\eta$ and $\beta$, as defined in Fig.~\ref{fig:system}, will be considered as the main tuning mechanism of the symmetry properties. 

\subsection{Crossing of orthogonal leaky modes}
The formation of EPs in our system hinges on the engineered crossing of orthogonal leaky modes that exhibit distinct radiative losses, as outlined in Section~\ref{sec:concept}. In this section, we demonstrate numerically that such mode crossings can be achieved by reducing the unit cell symmetry from $C_{4v}$ to $C_{2v}$.

We begin by analyzing how the photonic band dispersion within the light cone evolves with symmetry reduction. These results, obtained using the S$^4$ solver, are shown in Fig.~\ref{fig:dispersions_S4}. In particular, Fig.~\ref{fig:dispersions_S4}a shows the quasi-guided mode dispersion near the $\Gamma$ point ($k_x = k_y = 0$) for a photonic crystal slab with a square lattice ($\beta = 90^\circ$). The dispersion can be extracted by tracing the peaks in the absorption spectrum as a function of dimensionless frequency and in-plane wave vector. In this symmetric configuration, the $k_x$ and $k_y$ directions are equivalent, both corresponding to the high-symmetry $\Gamma$M direction in reciprocal space~\cite{Sakoda2005}. Consequently, the points M and $\bar{\mathrm{M}}$ are equivalent, and the band structure exhibits perfect symmetry along these two orthogonal directions, as expected for a $C_{4v}$-symmetric system.

A zoom-in of the first four band crossing at $\Gamma$ is shown in Fig.~\ref{fig:dispersions_S4}b, which clearly evidences the presence of a quadratic degeneracy in this high symmetry point corresponding to two bright modes, and two BICs, or dark modes (identified by vanishing signal at $k_x=k_y=0$), respectively. To highlight the different symmetries of the four modes, we explicitly plot the $z$-component of the electric field of these modes at the $\Gamma$ in Fig.~\ref{fig:dispersions_S4}c. By labeling the modes as $X_{ij}$, with $i,j=\{\sigma_x=\pm,\sigma_y=\pm\}$, we directly associate the bright (i.e., $X=B$) or dark ($X=D$) character of the corresponding mode at $\Gamma$ with their  symmetry with respect to the two relevant operations, $\sigma_x$ and $\sigma_y$: the two dark modes are both even (i.e., $i,j=+,+$) or odd (i.e., $i,j=-,-$) for either $\sigma_x$ or $\sigma_y$, while for bright modes the two symmetries are opposite.

The degeneracy of the two bright modes at $\Gamma$ is lifted by applying a structural distortion of the unit cell, as shown in Fig.~\ref{fig:dispersions_S4_zoom}b, in which the rhombic cell is defined by the angle $\beta=44^{\circ}$. The close up of Fig.~\ref{fig:dispersions_S4}b is  reported again in Fig.~\ref{fig:dispersions_S4_zoom}a, for completeness and direct comparison with the broken symmetry cases in Fig.~\ref{fig:dispersions_S4_zoom}b,c. For $\beta\neq 90^{\circ}$, the corresponding Brillouin zone becomes an irregular hexagon, for which the $k_x$ and $k_y$ directions are no longer equivalent (see top panels in Fig.~\ref{fig:dispersions_S4_zoom}b,c): the new high symmetry points along $k_x$ and $k_y$ are then defined as M$^\prime$ and K$^\prime$, respectively, reminding that the particular case of a perfectly hexagonal Brillouin zone corresponds to a triangular lattice in direct space, for which the corresponding high symmetry points are usually defined as M and K~\cite{Molding_the_flow,Sakoda2005}. The quasi-guided mode dispersion is then calculated along $\Gamma$M$^\prime$ and $\Gamma$K$^\prime$ high-symmetry directions, respectively. As shown in Fig.~\ref{fig:dispersions_S4_zoom}b, the dispersion along the two directions now becomes strongly asymmetric, and the quadratic degeneracy that is present at $\Gamma$ for $\beta=90^{\circ}$ has evolved into a Dirac-like degeneracy (i.e., a linear dispersion bands crossing) at a finite wave vector $k_x$ along $\Gamma$M$^\prime$ . This Dirac point remains quite close to $\Gamma$, in particular we have $k_{x,DP}\simeq 0.055 \pi/a$ and $k_y=0$. This crossing is due to the degeneracy of two orthogonal eigenmodes, indeed, as it can be inferred from the field profiles plotted in Figs.~\ref{fig:dispersions_S4}c. In Fig.~\ref{fig:dispersions_S4_zoom}c we also show the dispersion calculated for an elliptical hole whose long axis is aligned with the long axis of the rhombic unit cell. Hence, this further element does not change the lattice symmetry. However, the positions of bright and dark modes at $\Gamma$ is flipped, and the DP crossing at finite $k_x$ (which remains roughly in the same position) now comes from the degeneracy of two orthogonal modes, but having different symmetries as compared to the ones in Fig.~\ref{fig:dispersions_S4_zoom}b. We notice that the comparison between panels (a), (b), and (c) is made more consistent by assuming a constant filling fraction (i.e., the ratio between the hole and the unit cell areas is kept fixed).

In the following of the manuscript, we are going to focus our study on the complex energy dispersion of quasi-guided modes around this Dirac point, with the aim of systematically investigating the evolution of EPs and the related BFA as a function of the PhC structural parameters.

\begin{figure}[t]
    \centering
    \includegraphics[width=0.5\textwidth]{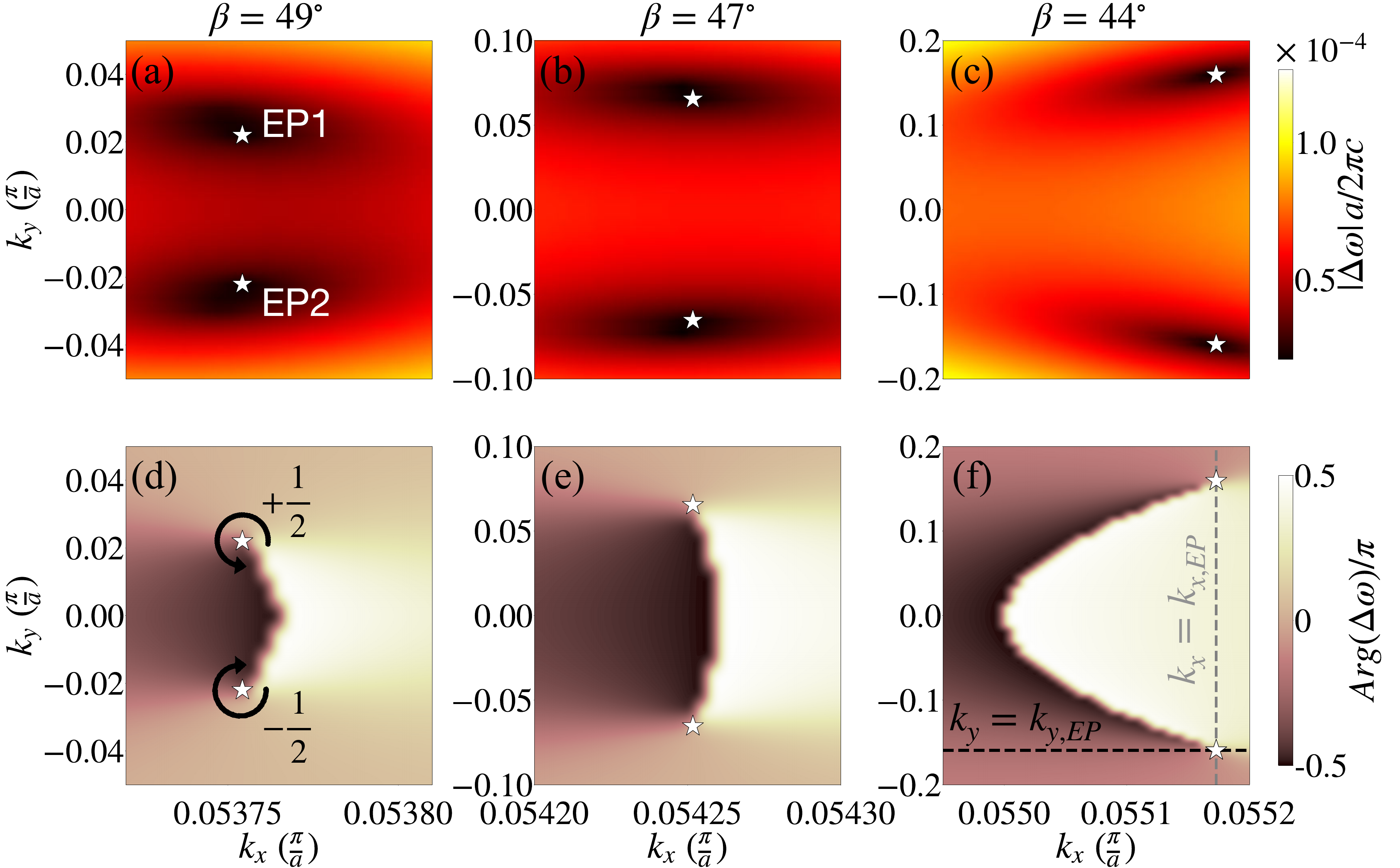}
    \caption{Complex energy difference between the two photonic branches around the Dirac point, calculated from COMSOL as a function of $k_x$ and $k_y$, for three different distorted unit cells, respectively showing the modulus with (a) $\beta=49^{\circ}$, (b) $\beta=47^{\circ}$, and (c) $\beta=44^{\circ}$, and the argument for (d) $\beta=49^{\circ}$, (e) $\beta=47^{\circ}$, and (f) $\beta=44^{\circ}$, clearly showing the evolution of the BFA at the degeneracy condition of the two eigenvalues. The EPs are clearly identified as two singular points symmetrically located with respect to the $k_y$ axis (e.g., at $(k_{x,EP}\sim 0.05517 \pi/a, k_{y,EP}\sim \pm 0.16 \pi/a)$ for $\beta=44^{\circ}$, see red markers). In all cases, The DP is clearly identified as the point on the BFA corresponding to $k_y=0$. The topological charges corresponding to the phase winding numbers around the EPs are also indicated. }
    \label{fig:Phase_beta}
\end{figure}

\begin{figure}[t]
    \centering
    \includegraphics[width=1\linewidth]{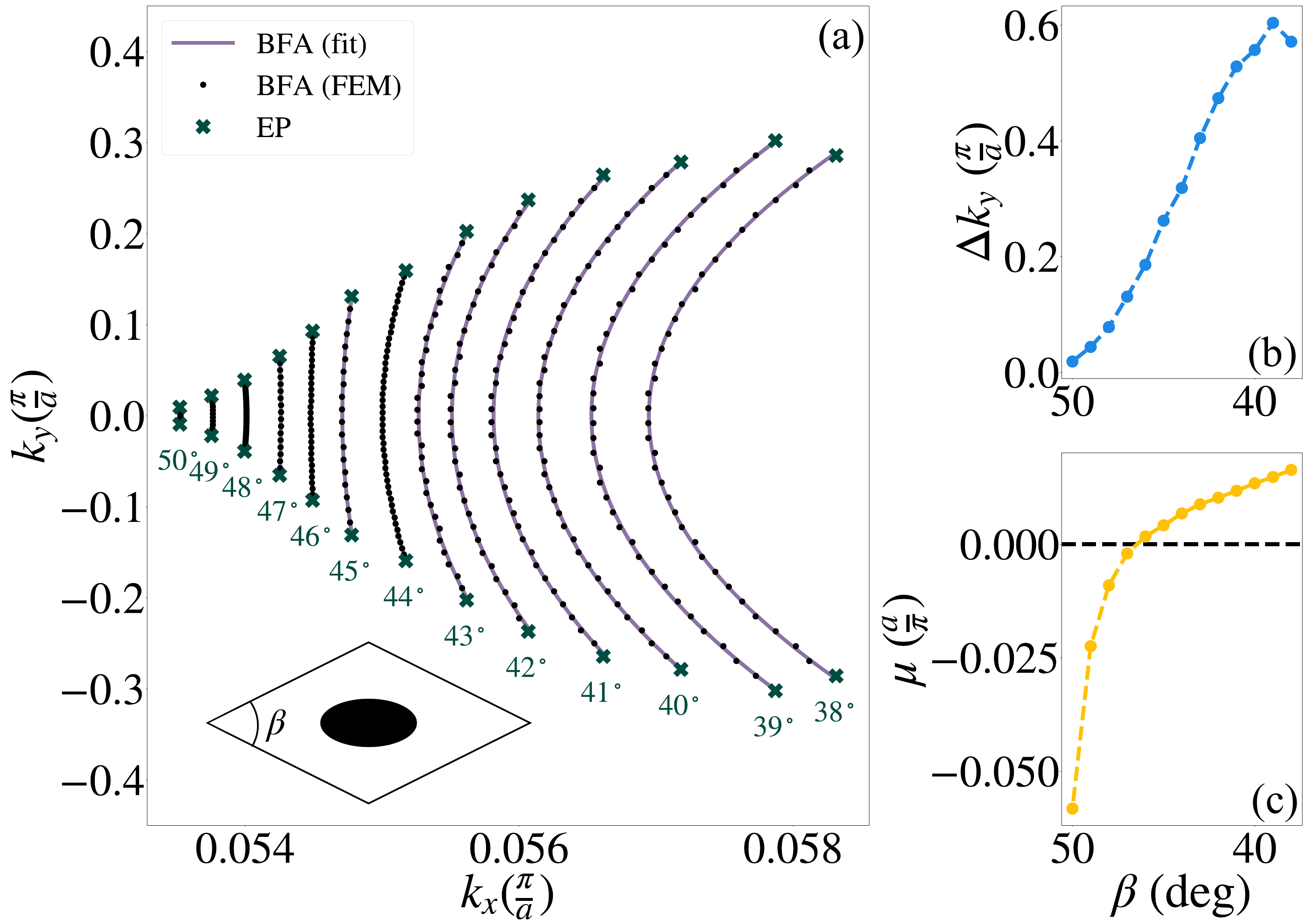}
    \caption{(a) Evolution of EPs and BFAs in the $k_x,k_y$ plane, as a function of the unit cell structural distortion parametrized by the angle $\beta$ ranging from $50^{\circ}$ to $38^{\circ}$ (points are extracted from COMSOL simulations by imposing the condition $\mathrm{Re}(\Delta\omega)=0$, and full lines are parabolic fits). The evolution is quantitatively summarized by plotting (b) the distance in $k_y$ of the EPs, as a function of $\beta$, and (c) the parabolic curvature fitted from the $k_x$-symmetric arcs (i.e., extracted from the full lines in panel a). }
    \label{fig:changing_beta}
\end{figure}

\subsection{Moving EPs and Shaping BFA}\label{sec:results}
\subsubsection{The $\eta=0^{\circ}$ case}
With the aim of investigating the detailed evolution of EPs in these systems, here we report the results of COMSOL simulations in the energy window around the Dirac point of Fig.~\ref{fig:dispersions_S4_zoom}c, by analyzing the complex energy dispersion in both $k_x$ and $k_y$ directions. 

Here, it is worth stressing that FEM is the only method allowing to obtain the complex dispersion around EPs with sufficient detail and accuracy, as compared to alternative approaches either in frequency or time domains.

The EPs and Fermi arc can be immediately visualized by plotting the complex energy gap between the two photonic branches  crossing at the DP at $(k_x=k_{x,DP},k_y=0)$, as a function of both $k_x$ and $k_y$ around this singularity. 
This quantity is calculated by running a great number of COMSOL simulations for the model parameters of Fig.~\ref{fig:dispersions_S4_zoom}c (i.e., $\beta=44^{\circ}$ and elliptical hole), for varying $(k_x , k_y)$ around the DP condition; from these simulations, two complex eigenenergies are numerically calculated, from which we straightforwardly get the complex energy difference $\Delta\omega$ between the two branches. We then extract the amplitude $|\Delta\omega|$ and the argument $\Phi=\mathrm{Arg}\{\Delta\omega\}$ of the complex gap. These latter quantities are then reported in Fig.~\ref{fig:Phase_beta}, for different values of $\beta$. From these results, we can clearly identify the two EPs, corresponding to the vanishing of $|\Delta\omega|$, (see, e.g., the first row in Fig.~\ref{fig:Phase_beta}, or equivalently as vortices in the texture of $\Phi$ (displayed in the second row of Fig.~\ref{fig:Phase_beta}), showing how each EP actually possesses a topological half-charge, as defined from the so called ``winding number'' \cite{Leykam2017}
\begin{equation}
    w = \frac{1}{2\pi}\oint_{\mathcal{C}}\nabla_{\mathbf{k}} \mathrm{Arg}\{\Delta\omega(\mathbf{k})\} \cdot d\mathbf{k} \, . 
\end{equation}
This non trivial topology is directly linked to the $\pi$ discontinuity at the BFA, which allows to nicely evidence the curve connecting the two EPs. In fact, both these properties are evident from the second row in Fig.~\ref{fig:Phase_beta}. 
As a consequence, the DP can be identified from the crossing of the Fermi arc with the $k_x$ axis, i.e., occurring for $k_y=0$ on the BFA. In addition, the DP degeneracy condition clearly evolves at $k_y \neq 0$ into a degeneracy of both real and imaginary parts, which marks the EP (see white star markers in Fig.~\ref{fig:Phase_beta}). 
It is also worth noting the different scale on the $k_x$ and $k_y$ axis: the EPs are symmetrically located with respect to $k_y=0$, they are separated by $\Delta k_y > 0.3 \pi/a$ for $\beta=44^{\circ}$, while they are basically at the same $k_x \simeq k_{x,DP}$, or, alternatively said, the curvature of the arc is quite negligible on the $\pi/a$ scale in momentum space. The evolution of the BFA is explicitly shown by plotting the same numerically calculated quantity corresponding to unit cell distortion angles $\beta=49^{\circ}$ in Fig.~\ref{fig:Phase_beta}a, and $\beta=47^{\circ}$ in  Fig.~\ref{fig:Phase_beta}b. In these cases, the distance between the EPs along $k_y$ is sizably reduced as compared to the case in Fig.~\ref{fig:Phase_beta}c, and the curvature changes sign. Interestingly, we also notice that the topological charge, associated to the winding number of the phase at the DP (which is zero), is split into two half-integers charges with opposite signs at the EPs, as explicitly indicated in the Figure.  

\begin{figure}[t]
    \centering
    \includegraphics[width=1\linewidth]{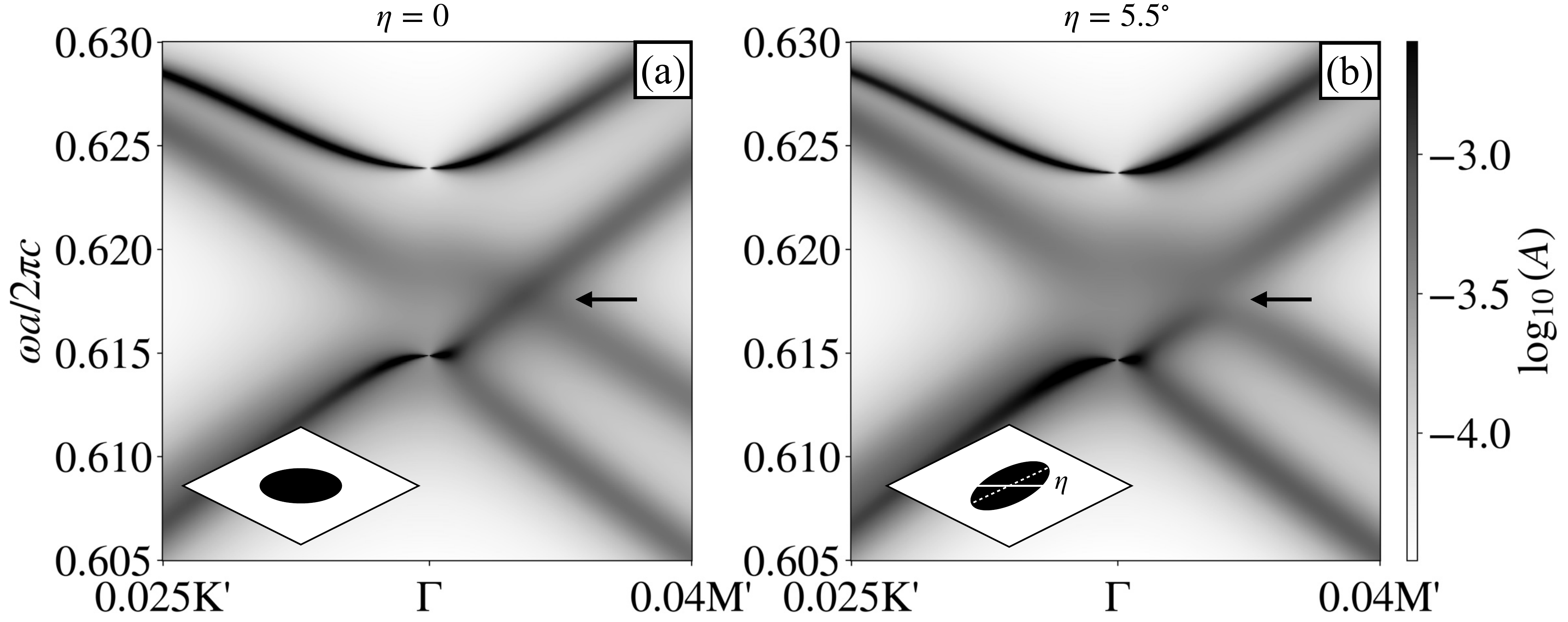}
    \caption{Dispersion within the light cone in the vicinity of the $\Gamma$ point and along two orthogonal directions ($x$ and $y$), for the distorted lattice structure with $\beta=80^{\circ}$, and for (a) $\eta=0^{\circ}$, (b) $\eta=5.5^{\circ}$, respectively.
    Numerical simulations of the absorption spectra have been obtained by RCWA, and quasi-guided modes are clearly visible as peaks in the variable angle absorption spectra (log scale, log$_{10}(A)$). The arrows indicate the position of the DP at $\eta=0^{\circ}$, whose degeneracy along $x$ is lifted for $\eta=5.5^{\circ}$. Photonic crystal parameters are as in Fig.~\ref{fig:dispersions_S4}, and all the energies are then given in dimensionless units, $a/\lambda=\omega a/2\pi c$.}
    \label{fig:bands_changing_eta}
\end{figure}

To better visualize the evolution of the BFA, we report the positions of EPs and BFAs as a function of $\beta$ in Fig.~\ref{fig:changing_beta}, for several values of this structural parameter. In fact, for a useful visualization of the different conditions within the same plot, we report in Fig.~\ref{fig:changing_beta}a the points corresponding to the degeneracy condition along the Fermi arc, i.e., $\mathrm{Re}(\Delta\omega)=0$, as obtained from the COMSOL simulations for different values of $\beta$ ranging from $50^{\circ}$ to $38^{\circ}$ in single degree steps. As it can be seen, a little variation in $\beta$ produces a strong modulation of the corresponding BFA, as well as the distance between the two EPs along $k_y$. In particular, in Fig.~\ref{fig:changing_beta}b we report the vertical distance between the EPs, i.e., $\Delta k_y$, showing that it can be controlled by more than an order of magnitude, from fractions of $\pi/a$ to about $0.6\pi/a$, i.e., a significant extension within the Brillouin zone. In addition, to quantitatively assess the changes in the BFA we also extract the parabolic curvature $\mu$, defined by fitting each Fermi arc with the curve $k_x=k_C + \mu k_y^2$. The corresponding results are reported in Fig.~\ref{fig:changing_beta}c as a function of $\beta$, clearly showing that within a few degrees in structural distortion of the unit cell the Fermi arc curvature can be modulated by orders of magnitude, and even go from negative to positive. For some $\beta$ between $47^{\circ}$ and $46^{\circ}$ the Fermi arc actually becomes a vertical straight line (i.e., $\mu=0$). 

\begin{figure}[t]
    \centering
    \includegraphics[width=1\linewidth]{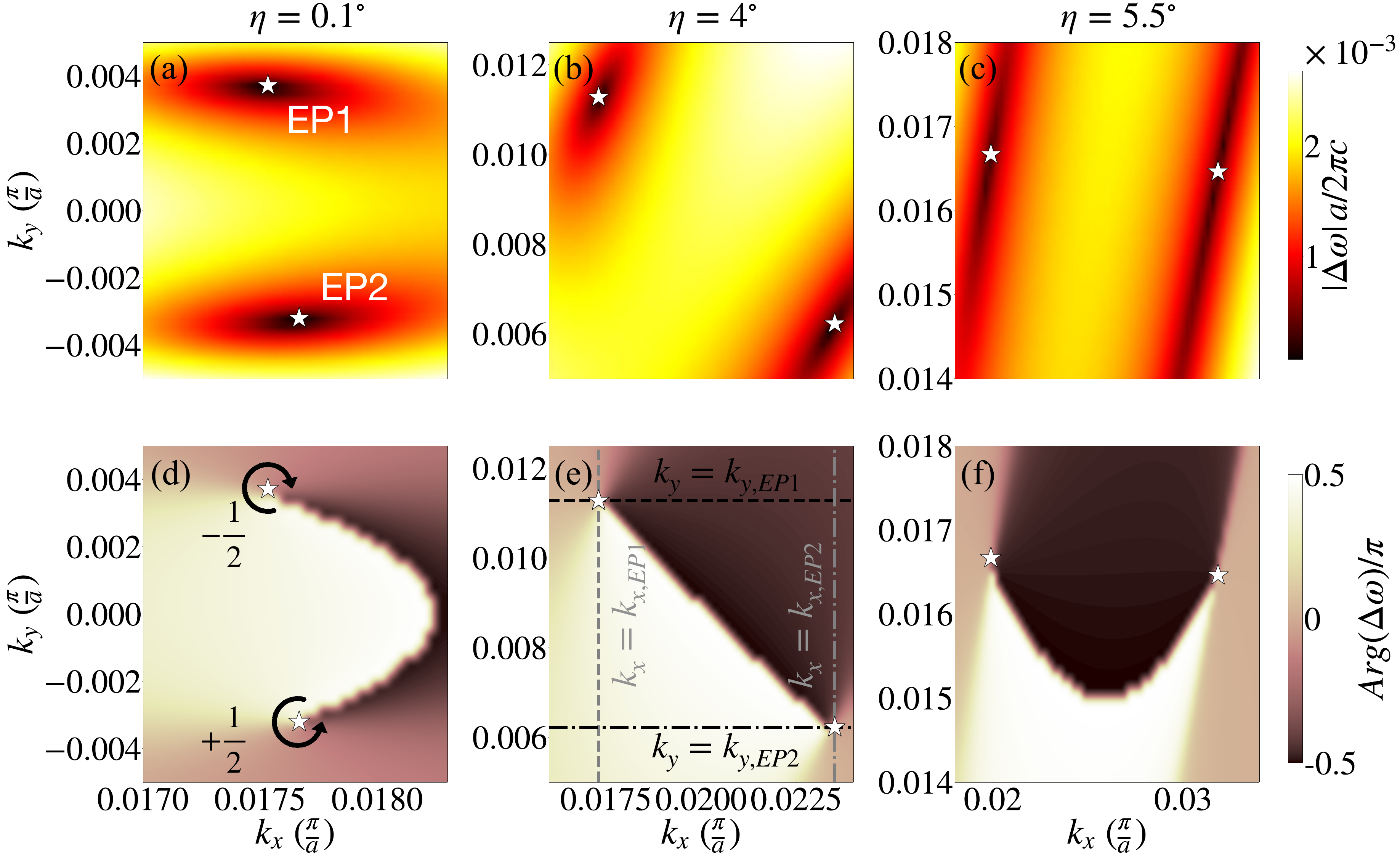}
    \caption{Evolution of the Fermi arc and EPs in the two-dimensional reciprocal plane, directly visualized from the argument of the complex energy gap numerically calculated from COMSOL, as a function of the tilting angle of the elliptical hole with respect to the horizontal axis, at fixed $\beta=80^{\circ}$, and for (a) $\eta=0.1^{\circ}$, (b) $\eta=4^{\circ}$, and (c) $\eta=5.5^{\circ}$.}
    \label{fig:Phase_eta}
\end{figure}

\subsubsection{The $\eta\neq0^{\circ}$ case}

So far, we have studied the evolution of EPs and BFAs by keeping the symmetry with respect to the vertical  $(x,z)$ plane within the elementary cell, i.e., breaking the $C_{4v}$ symmetry but not the $C_{2v}$. As a consequence, the resulting EPs are symmetrically located with respect to $k_y=0$, as clearly seen in Fig.~\ref{fig:changing_beta}. As already mentioned before, we now apply a tilting of the elliptical hole with respect to the horizontal axis by the angle $\eta$ (see Fig.~\ref{fig:system}), which breaks also the $C_{2v}$ symmetry, and produces a non-intuitive evolution of both EPs and BFAs. 

First, we explicitly show in Fig.~\ref{fig:bands_changing_eta} how the band structure calculated by RCWA evolves around the $\Gamma$ point and along the two orthogonal directions (as already done, e.g., in Fig.~\ref{fig:dispersions_S4_zoom}). We report results for $\eta=0^{\circ}$ and $\eta=5.5^{\circ}$, after fixing $\beta=80^{\circ}$. We choose this condition since it allows to better visualize the evolution of Fermi arcs and DP as a function of $\eta$. In fact, the DP degeneracy along the $x$ direction is now lifted, as it is evident in Fig.~\ref{fig:bands_changing_eta}b (see  the pointing arrows). 

\begin{figure}[t]
    \centering
    \includegraphics[width=1\linewidth]{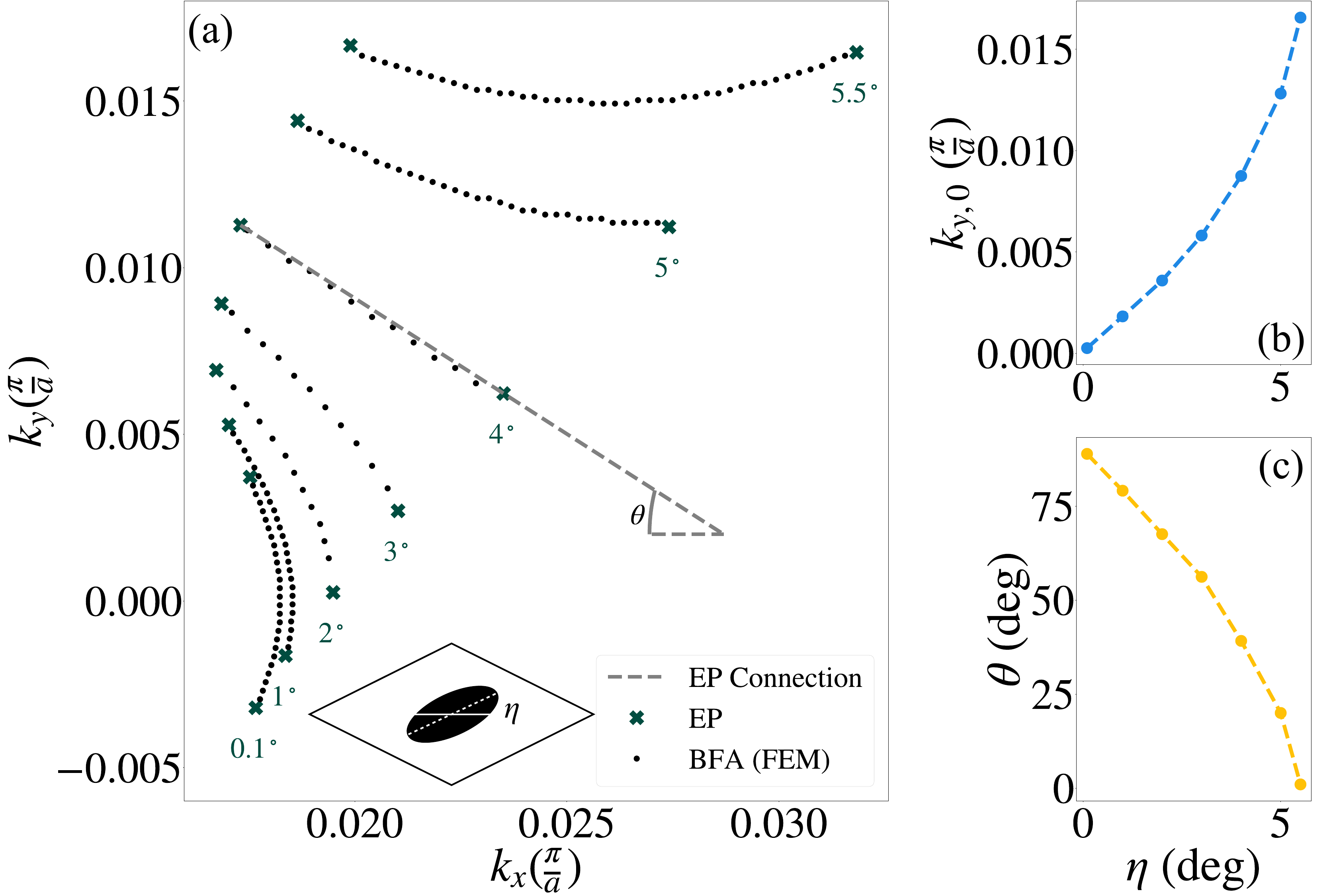}
    \caption{(a) Evolution of the EPs and the BFA 
    in the $k_x,k_y$ plane, for different values of $\eta$, from the data calculated with COMSOL and imposing the condition $\mathrm{Re}(\Delta\omega)=0$. Further relevant information can be gathered by looking at (b) the middle point along the BFA ($k_{y,0}$) and (c) the tilting of the direction identified by the line connecting the two EPs with respect to horizontal direction, showing that $\theta$ can be controlled to essentially go from $90^{\circ}$ to $0^{\circ}$ upon increasing $\eta$. }
    \label{fig:changing_eta}
\end{figure}

On the other hand, the whole evolution of the band structure should be considered, in both $k_x$ and $k_y$. 
So, to give a visual idea of how the DP, EPs, and BFA evolve in each case, we report in Fig.~\ref{fig:Phase_eta} the calculated complex energy gap between the two relevant eigenmodes. In particular, we explicitly show the modulus and argument of the complex gap in the $(k_x,k_y)$ plane for different values of $\eta$. With respect to the results presented in the previous Figures, here we are closer to $\Gamma$ along the $k_x$ direction. In fact, the DP appears around $k_x\simeq 0.018 \pi/a$ when $\eta=0.1^{\circ}$ (Fig.~\ref{fig:Phase_eta}a), and the two EPs are still quite symmetrically located with respect to $k_y=0$. However, by allowing for a stronger tilting of the elliptical hole orientation, the EPs rapidly evolve along the Fermi arc, and at $\eta=4^{\circ}$ (Fig.~\ref{fig:Phase_eta}f) there is actually no DP along $k_x$ any more, while the two EPs are both located in the upper plane within the Brillouin zone. In the band structure, this condition appears as a lifting of the crossing modes degeneracy along $k_x$ (as in Fig.~\ref{fig:bands_changing_eta}b, for example). Interestingly, at $\eta=4^{\circ}$ we can also appreciate a change in the Fermi arc curvature, which is even more evident in Fig.~\ref{fig:Phase_eta}f. Another interesting point to be highlighted is that there appears to be a flipping of the topological charge sign between the upper and lower half planes, as compared to the $\beta$ values around 40$^\circ$ (see, for example, a direct comparison between Figs.~\ref{fig:Phase_beta}d and \ref{fig:Phase_eta}d). The latter point might deserve further theoretical investigation. \\
Finally, the full evolution of the EPs and BFAs in the $(k_x,k_y)$ plane as a function of $\eta$ is displayed in Fig.~\ref{fig:changing_eta}, in the range from $\eta=0.1^{\circ}$ to $5.5^{\circ}$, directly reporting the eigenvalues outcomes of COMSOL. First, we report the evolution obtained as before by imposing the usual BFA condition, $\mathrm{Re}(\Delta\omega)=0$, for different values of $\eta$. The result is then visualized in  Fig.~\ref{fig:changing_eta}a. Interestingly from this plot, the BFA evolves from an almost parabolic arc with positive curvature and symmetric with respect to $k_y=0$ at $\eta=0^{\circ}$, to a negative curvature one, and all the way to an almost parabolic arc with symmetry axis orthogonal to $k_x$. This can be best appreciated by looking at the mid-point between the two EPs ($k_{y,0}$), as well as the angle between the direction identified by the line connecting the two EPs and the horizontal (i.e., $\hat{\mathbf{x}}$) direction, $\theta$. Both quantities are plotted as functions of $\eta$ in Figs.~\ref{fig:changing_eta}b,c. Strikingly, $\theta$ can be controlled to range from $90^{\circ}$, i.e., having the parabola symmetry axis aligned with $k_x$, to almost $0^{\circ}$, i.e., the symmetry axis of the parabolic curve aligned with $k_y$ (see Fig.~\ref{fig:changing_eta}c).

In summary, the results shown in this Section highlight the high tunability of EPs and BFAs within the 2D Brillouin zone of a PhC slab with rhombic unit cell and an elliptical hole basis, in which only two structural parameters are used to continuously break either $C_{4v}$ or $C_{2v}$ symmetries. In particular, these recipes allow to tailor the position of the two EPs, and the extension as well as the orientation and curvature of the BFA, thus providing an interesting playground to address non-Hermitian physics in a practical and cost-effective dielectric platform.

\section{\label{conclusion} Conclusions}

We have performed systematic numerical simulations to study the evolution of Dirac points, exceptional points, and bulk Fermi arcs in purely dielectric two-dimensional photonic crystal slabs, i.e., thin semiconductor or insulating films patterned with a lattice of elliptical holes. 
While several studies have characterized the position of exceptional points in momentum space, little has been published regarding the engineering of the bulk Fermi arcs connecting a pair of exceptional points from structural properties of the underlying photonic crystal lattice. In particular, we have investigated in detail how to employ a reduction of lattice symmetry from $C_{4v}$ to $C_{2v}$ to first generate a Dirac point at finite momentum (here, $k_x \neq 0$) out of a quadratic degeneracy at $k_x=0=k_y$. Exceptional points appear then at $k_x\neq 0$ and $k_y\neq 0$, connected by the bulk Fermi arc, the region of momentum space defined by the degeneracy of the real part of the complex energy eigenvalues. The curvature of the Fermi arc can then be tailored by simply tuning the lattice cell distortion from the original square lattice. An additional tuning knob is then given by the orientation of the central elliptical hole with respect to the coordinate axes, which allows to further tune not only the curvature of the Fermi arc and the exceptional points distance in the reciprocal plane, but also the orientation of the arc itself, possibly rotating by $90^{\circ}$ with respect to the original alignment. 

Beyond the specific photonic crystal slab design investigated here, our work opens compelling perspectives for future studies and practical applications in the broader context of non-Hermitian topology and photonics. In particular, the general approach of engineering EPs through controlled symmetry breaking is not restricted to the rhombic lattice explored here, but can be readily extended to a variety of photonic crystal structures. For instance, simpler geometries such as one-dimensional photonic gratings~\cite{HaiSon_Unveiling}, or alternative two-dimensional lattices (e.g., the hexagonal~\cite{Kang2022}, or the monoclinic~\cite{Toftul2024} lattices), as well as more complex architectures such as bilayer photonic crystals~\cite{Letartre2022,Ni2024} and moiré photonic structures~\cite{lu2020,Tang2021,Nguyen2022}, could similarly exploit symmetry-driven control to realize tailored EPs. These extended platforms offer exciting opportunities to investigate advanced phenomena, including geometry-dependent skin effects~\cite{Zhang2022,Fang2022,Zhou2023}, where wave localization becomes strongly dependent on system boundaries and symmetries. Additionally, EPs emerging from tailored symmetry breaking provide a versatile platform to investigate non-Hermitian band topology beyond conventional Dirac or Weyl points~\cite{Kunst_Review}. Such studies promise to enrich both fundamental insights and device-oriented innovations in nanophotonics and integrated optics.

\begin{acknowledgments}
The authors thank the NanoLyon platform with its computation facilities. L.F. acknowledges the ``Erasmus+ traineeship'' program from the University of Pavia for partly funding his eight months stay at the \'Ecole Centrale de Lyon under the supervision of H.S.N. This work was partially supported by the French National Research Agency (ANR) under the project POLAROID (ANR-24-CE24-7616-01). 
The authors warmly acknowledge Lucio C. Andreani, Taha Benyattou, and Xavier Letartre for several useful and inspiring discussions during the preliminary stages of this work.
\end{acknowledgments}

\bibliography{EP_FA_biblio}

\onecolumngrid
\appendix

\section{Additional plots}

\begin{figure}[b]
    \centering
    \includegraphics[width=0.8\linewidth]{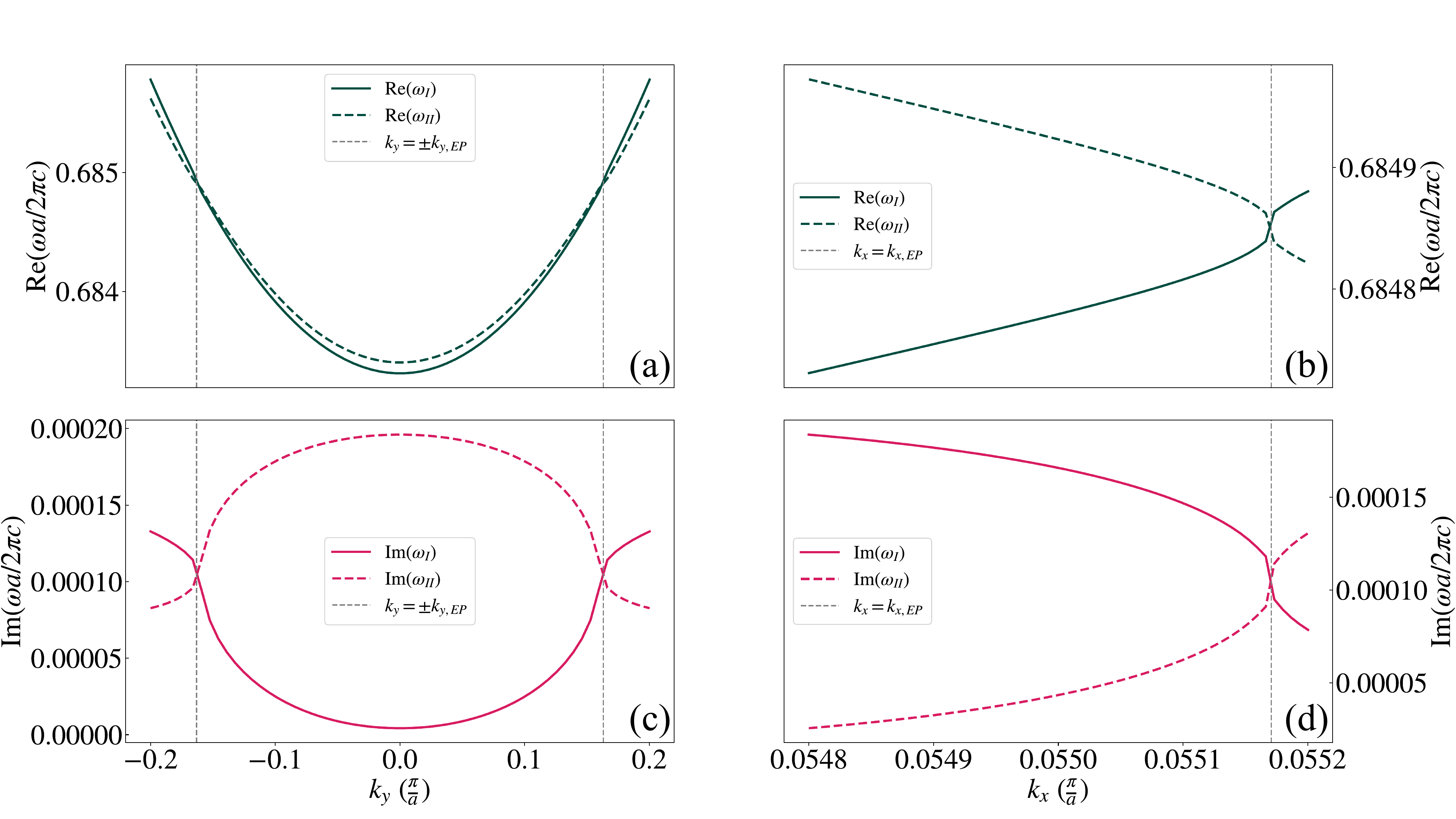}
    \caption{Complex energy dispersion, resolved separately as a function of $k_x$ and $k_y$, as obtained from  COMSOL simulations: (a) real and (c) imaginary parts of eigenenergies of the two modes, calculated at $k_x \sim 0.05517 \pi/a$ as a function of $k_y$; the positions of the EPs are marked by vertical dashed lines in both panels (highlighting the degeneracy of both real and imaginary parts). These plots correspond the the cut highlighted as a vertical dashed line in Fig.~4f. Correspondingly, we also report the (b) real and (d) imaginary parts of the eigenenergies calculated at $k_y \sim 0.16\pi/a$ and as a function of $k_x$, which relates to the horizontal dashed line in superimposed to the plot in Fig.~4f. Again, the EP is marked by a vertical dashed lines in both plots. }
    \label{fig:dispersions_COMSOL}
\end{figure}

To further highlight the actual presence of EP along the Fermi arc, we plot the two eigenvalues, $\omega_{+}$ and $\omega_{-}$, separately for real and imaginary parts, in correspondence of the two cuts evidenced by dashed lines in Fig.~\ref{fig:Phase_beta}, i.e., as a function of $k_y$ and fixed $k_{x}=k_{x,EP}$ and as a function of $k_x$ and fixed $k_{y}=k_{y,EP}$. The results are reported in Fig.~\ref{fig:dispersions_COMSOL}.

\begin{figure}[t]
    \centering
    \includegraphics[width=0.8\linewidth]{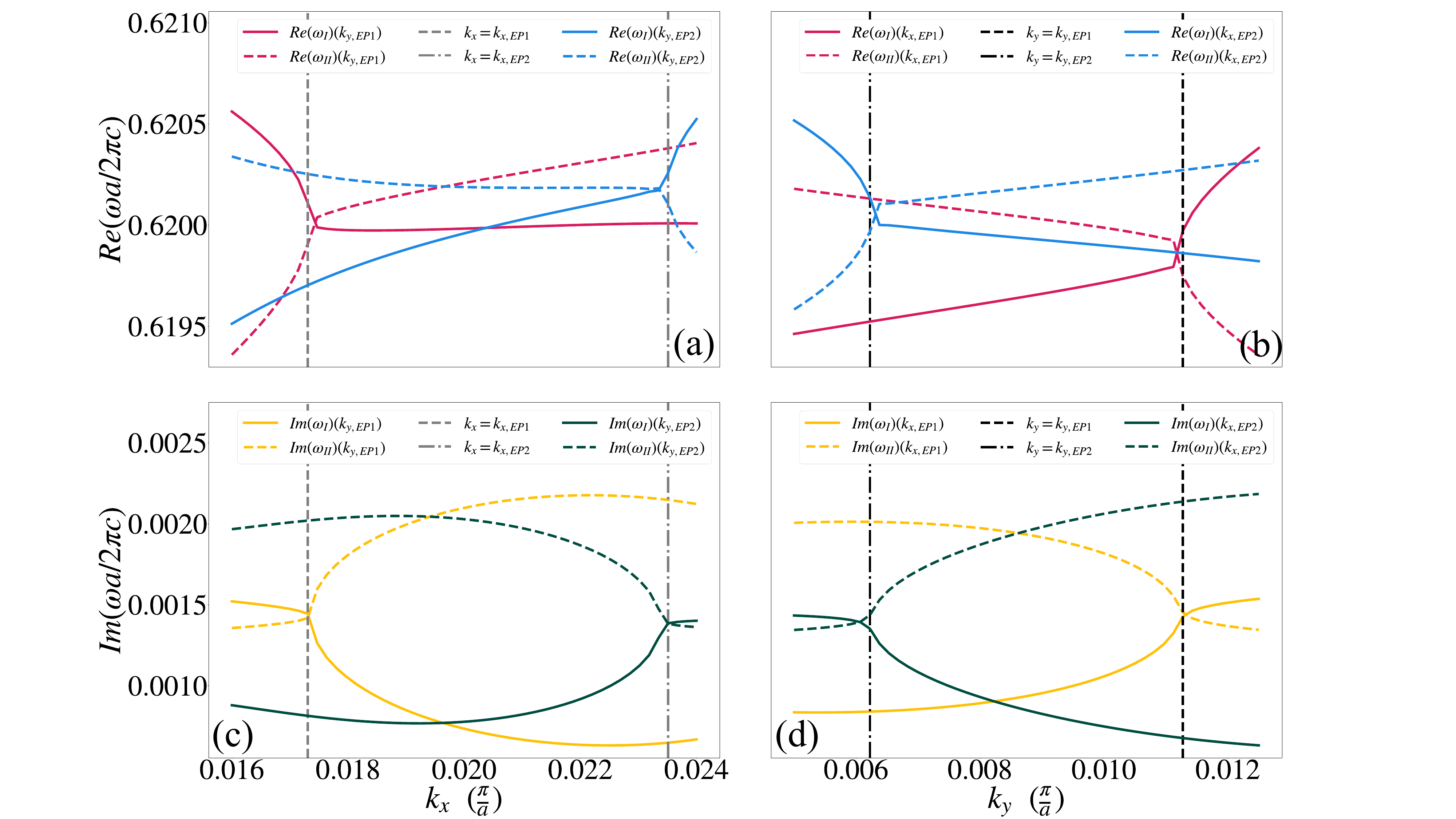}
    \caption{Selected cuts from the complex dispersion of the two photonic branches which give rise to the plot of Fig.~\ref{fig:Phase_eta}c, i.e., for $\eta=4^{\circ}$: (a) real and (c) imaginary parts of the two eigenvalues as a function of $k_x$ and at fixed $k_y$ values corresponding to the crossing of the two EPs (see horizontal dashed lines in Fig.~\ref{fig:Phase_eta}e); correspondingly, we also plot the (b) real and (d) imaginary parts of the two branches eigenvalues as a function of $k_y$ for fixed $k_x$ values corresponding to the two vertical dashed lines in Fig.~\ref{fig:Phase_eta}e. In all the panels, the related coordinates in correspondence to the EPs are evidenced by vertical dashed lines. }
    \label{fig:dispersions_COMSOL_eta=4}
\end{figure}

In addition, to better appreciate the formation of EPs after $C_{2v}$ symmetry breaking, we report in Fig.~\ref{fig:dispersions_COMSOL_eta=4} the real and imaginary parts of the complex $\omega_{\pm}$ eigenvalues obtained from the COMSOL simulations, both as a function of $k_x$ and at fixed $k_y$ (fixed by the crossing of EPs), and as a function of $k_y$ and at fixed $k_x$. These four possibilities correspond to the two vertical and horizontal line-cuts in Fig.~\ref{fig:Phase_eta}c. The vertical lines in the plots identify the EPs, as the condition for which both real and imaginary parts of the two branches become degenerate. However, due to the discretization in the input parameters, the degeneracy at a given $k_x$ or $k_y$ point does not precisely coincide between real and imaginary eigenvalues. The physical conclusions that can be drawn by the results presented here are not affected by this slight discrepancy.

\end{document}